\documentclass[smallextended]{svjour3}     
\smartqed  
\usepackage{graphicx}
\newcommand{\Msol}{M_\odot}

\newcommand{\degree}{^\circ}

\begin{document}

\def\Msol{M_\odot}
\newcommand{\ud}{{\mathrm d}}
\title{Microlensing as a probe of the Galactic structure}

\subtitle{20 years of microlensing optical depth studies}


\author{
Marc Moniez
}


\institute{M. Moniez \at
        Laboratoire de l'Acc\'{e}l\'{e}rateur Lin\'{e}aire,
IN2P3 CNRS, Universit\'e Paris-Sud, 91405 Orsay Cedex, France\\
              Tel.: +33-1-64468344\\
              Fax: +33-1-64468397\\
              \email{moniez@lal.in2p3.fr}           
}

\date{Received: date / Accepted: date}
\maketitle
\begin{abstract}
Microlensing is now a very popular observational astronomical
technique.
The investigations accessible through this effect range from
the dark matter problem to the search for extra-solar
planets.
In this review, the techniques to search for microlensing
effects and to determine optical depths
through the monitoring of large samples of stars will
be described.
The consequences of the published results on the knowledge
of the Milky-Way structure and its dark matter component
will be discussed.
The difficulties and limitations of the ongoing programs and the
perspectives of the microlensing optical depth technique
as a probe of the Galaxy structure will also be detailed.
\end{abstract}
\section{Introduction}
In 1936 A. Einstein published a half-page note in Science entitled
{\it ``lens-like action of a star by the deviation of
light in the gravitational field''} \cite{Einstein}, with this final
comment: {\it ``There is not great chance of observing this phenomenon.''}
According to J. Renn \cite{Renn} the effect was already predicted by A. Einstein as early
as 1912, before completion of the general theory of relativity.
S. Liebes \cite{Liebes} considered again the properties of the gravitational lenses
in 1964, and studied the possibility to detect high magnification events.
Furthermore, this author already mentioned the possibility to detect
invisible compact objects through microlensing.
In 1986, B. Paczy\'nski \cite{pacz1986} pointed out the possibility of using
the gravitational microlensing effect to detect massive compact objects of
the Galactic halo in the direction of the Magellanic Clouds.

In 1989, the raising question amongst astrophysicists was the nature
of dark matter within the Milky-Way halo.
WIMPS\footnote{Weakly Interacting Massive ParticleS}
search experiments were already looking for exotic dark matter, but
it was realized at this epoch that the question
of today's nature of the majority of the baryons (at $z=0$)
---still pending--- could be addressed through the microlensing
technique.
From 1989, several groups started survey programs to search for compact halo
objects within the Galactic halo. The challenge for the EROS and
MACHO teams was to clarify the
status of the missing hadrons in our own Galaxy.
In september 1993, three teams, EROS \cite{eroslmc},
MACHO \cite{machlmc} and
OGLE \cite{oglpr} discovered the first microlensing
events in the directions of the Large Magellanic Cloud and
the Galactic Center. 
Since these first discoveries, thousands of microlensing effects have been
detected in the direction of the Galactic Center (GC) together with
a handful of events towards the Galactic Spiral Arms (GSA) and the
Magellanic Clouds.
\vspace{-2mm}
\paragraph{\bf Probing Galactic structure with microlensing:}
Microlensing has proven to be a powerful probe of the
Milky-Way structure, and not only for the hidden baryonic matter.
Searches for microlensing towards the Magellanic Clouds (LMC, SMC)
and M31 provide optical depths through the Galactic halo,
allowing one to study dark matter under the form of massive compact objects.
Searches towards the Galactic plane (Galactic Center, GC and Galactic
Spiral Arms, GSA) allow one to measure the optical depth
due to ordinary stars in the Galactic disk and bulge.
Kinematical models as well as mass functions can also be tested through
the event duration distributions.
In a few special cases called ``exotic'' events,
more specific information on the lensing system
has been obtained; unfortunately they don't yet provide sufficiently
large samples to extract reliable information on the Galactic
structures.

This article reviews the optical depth measurements,
which are the instantaneous probabilities for a point-source of a target
to be magnified by a factor larger than $1.34$.
The description of the observational devices and of the analysis
techniques will be given, followed by a discussion about the main
difficulties and limitations which can be overcome in
future surveys.
The consequences of the optical depth measurements
on the Galactic structure knowledge will also be
reviewed, with the first connections that can be established
between the kinematics of the Galactic structures and the duration
distributions.
We will conclude with a prospective on the future surveys
and a description of the potential of microlensing searches
in infrared and of ``fine'' searches, sensitive to exotic events,
providing valuable extra information.

\section{Microlensing basics}
\label{sec:basics}
\begin{figure}
\begin{center}
\includegraphics[width=9cm]{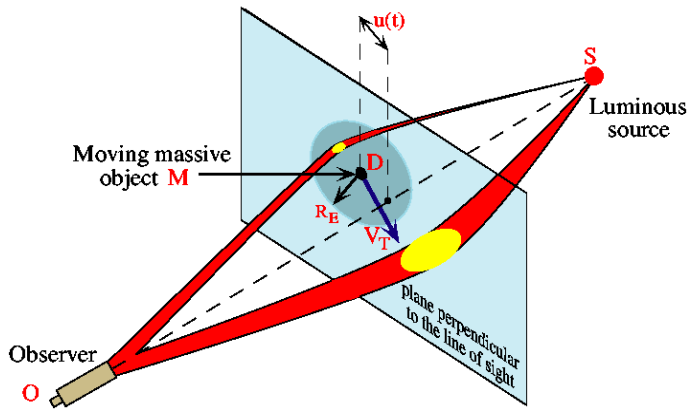}
\caption{\it Principle of the microlensing effect:
As the deflector D of mass M moves with a transverse relative speed
$V_T$, the impact parameter u(t) changes with time, and
so does the magnification of the source.
\label{miclens}}
\end{center}
\end{figure}

Gravitational microlensing effect occurs when a massive compact object
passes close to the line of sight of a star
and produces gravitational images that are not intercepted by
the massive object (no eclipse), but that cannot be separated
in telescopes (no multiple images)(see Fig. \ref{miclens}).
The detection of such a coincidence is then possible only if a
measurable magnification variation occurs during
the observation time.
%
The details on the formalism associated with this phenomenon
can be found in \cite{Schneider} and we will only provide here
a few basic tools. 
In the approximation of a single point-like object acting as a
deflector on a single point-like source,
the total magnification of the source luminosity
at a given time $t$ is the sum of the contributions of
two images, given by
\begin{equation}
\label{magnification}
A(t)=\frac{u(t)^2+2}{u(t)\sqrt{u(t)^2+4}}\ ,
\end{equation}
where $u(t)$ is the distance of the deflecting object
to the undeflected line of sight, expressed
in units of the ``Einstein Radius" $R_{\mathrm{E}}$:
\begin{eqnarray}
R_{\mathrm{E}} = \sqrt{\frac{4GM}{c^2}D_S x(1-x)}
\simeq 4.54\ A.U. \times\left[\frac{M}{\Msol}\right]^{\frac{1}{2}}
\left[\frac{D_S}{10\ kpc}\right]^{\frac{1}{2}}
\frac{\left[x(1-x)\right]^{\frac{1}{2}}}{0.5}.
\end{eqnarray}
Here $G$ is the Newtonian gravitational constant,
$D_S$ is the distance of the observer to the source and
$x D_S=D_L$ is its distance to the deflector of mass $M$.
The motion of the deflector relative to the line of sight
makes the magnification vary with time.
Assuming a deflector moving at a constant relative transverse
speed $v_T$, reaching its minimum
distance $u_0$ (impact parameter) to the undeflected line of sight
at time $t_0$, $u(t)$ is given by
\begin{equation}
\label{impact}
u(t)=\sqrt{u_0^2+((t-t_0)/t_{\mathrm{E}})^2},
\end{equation}
where $t_{\mathrm{E}}=R_{\mathrm{E}} /v_T$, the ``lensing time scale", is
the only measurable parameter
bringing useful information on the lens configuration in the
approximation of simple microlensing:
\begin{eqnarray}
t_{\mathrm{E}} \sim
79.\ days\left[\frac{v_T}{100\, km/s}\right]^{-1}
\left[\frac{M}{\Msol}\right]^{\frac{1}{2}}
\left[\frac{D_S}{10\, kpc}\right]^{\frac{1}{2}}
\frac{[x(1-x)]^{\frac{1}{2}}}{0.5}\; . 
\end{eqnarray}
\subsection{Microlensing event characteristics}
The so-called simple microlensing effect (point-like source and
point-like lens
with uniform relative motion with respect to the line of sight)
has some characteristic
features which allow one to discriminate it from any known
intrinsic stellar variability :
\begin{itemize}
\item The event is singular in the history of the source
(as well as of the deflector).
\item The gravitational origin of the effect implies that the
magnification is independent of the color.
\item The magnification is a known function of
time, depending on only 3 parameters ($u_{min}, t_0, t_{\mathrm{E}}$),
with a symmetrical shape.
\item As the geometric configuration of the source-deflector system
is random, the impact parameters
of the events must be uniformly distributed.
\item The passive role of the lensed stars implies that their
population should be representative of the monitored
sample, particularly with respect to the observed color and
magnitude distributions.
\end{itemize}

This simple microlensing description can be broken in
many different ways: the lens may be a double system \cite{mao1995},
or the source may be an extended object \cite{Yoo},
or the relative motion with respect to the line of sight may deviate
from a uniform motion due either to the rotation of the Earth around
the Sun (parallax effect) \cite{Gould92} \cite{Hardy95},
or to an orbital motion of the source or of the lens around
the center-of-mass of a multiple system 
\cite{Mollerach}.
All these deviations have been observed
and will be discussed further in this issue (see the contribution of M. Dominik \cite{Dominik2009}).
The interest of these ``exotics'' comes from the extra-information
on the lensing configuration that can be obtained in such specific cases.
\subsection{The observables: optical depth, event rate and $t_{\mathrm{E}}$ distribution}
The optical depth up to a given source distance $D_S$ is defined as the
instantaneous probability for the line of sight of a target source to intercept
a deflector's Einstein disk, that corresponds to a magnification $A > 1.34$.
Assuming that the distribution of the deflector masses is
described by a density function $\rho(D_L)$ and a
normalized mass function $\ud n_L(D_L,M)/\ud M$, this probability is:
\begin{equation}
\tau(D_S)=\int_0^{D_S}\int_{M=0}^{\infty} \pi \theta_E^2 \times
\frac{\rho(D_L)D_L^2}{M}\frac{\ud n_L(D_L,M)}{\ud M}\ud M \ud D_L\, ,
\end{equation}
where $\theta_E=R_{\mathrm{E}}/D_L$ is the angular Einstein radius of a lens of mass
$M$ located at $D_L$.
The second term of the integral is the differential number of these lenses
per mass unit and per solid angle.
As the solid angle of the Einstein disk is proportional to
the deflectors' mass $M$, this probability is found to be independent of
the deflectors' mass function:
\begin{equation}
\tau(D_S)=\frac{4 \pi G D_S^2}{c^2}\int_0^1 x(1-x)\rho(x) \ud x\, ,
\end{equation}
where $\rho(x)$ is the mass density of deflectors located at a distance $x D_S$.
The mean optical depth towards a given population
defined by a distance distribution $\ud n_S(D_S)/\ud D_S$
of target stars is defined as
\begin{equation}
<\tau>
=\frac{\int_0^{\infty} \frac{\ud n_S(D_S)}{\ud D_S} \tau(D_S)D_S^2\, \ud D_S}
{\int_0^{\infty} \frac{\ud n_S(D_S)}{\ud D_S}  D_S^2\, \ud D_S}.
\label{optpop}
\end{equation}

The {\it measured} optical depth associated to microlensing events
observed in a population of $N_{obs}$ stars monitored for a duration
$\Delta T_{obs}$ is given by
\footnote{The product $N_{obs}\times \Delta T_{obs}$
is called the exposure.}
\begin{equation}
\tau =\frac{1}{N_{obs}\Delta T_{obs}}\frac{\pi}{2}\sum_{events}
\frac{t_{\mathrm{E}}}{\epsilon (t_{\mathrm{E}})},
\label{taumeas}
\end{equation}
where 
$\epsilon (t_{\mathrm{E}})$ is the average detection efficiency
of microlensing events with a time scale $t_{\mathrm{E}}$ (see the precise
definition in Sect. \ref{sec:efficiency}).

The calculated optical depth does not depend on the deflectors' mass
function, but the measured optical depth, that takes into account
the mean detection efficiency $\epsilon (t_{\mathrm{E}})$, can be biased
by this mass function, in particular because the detection function
vanishes for very short or very long duration events.
This fact makes
impossible a perfect compensation of the inefficiencies.
If many deflectors are light (resp. heavy) enough to produce extremely
short (resp. long) duration events that cannot be detected, the
measured optical depth will clearly be underestimated.
This is why collaborations often indicate that their results
are valid within a given duration domain.
The control of the detection efficiency ---that is needed
to get reliable optical depths and rate estimates---
is one of the most delicate aspects of the survey searches.
\vspace{-2mm}
\paragraph{\bf Event rate:}
Contrary to the optical depth, the microlensing event rate
depends on the deflectors' mass distribution as well as on the
velocity and spatial distributions.
The global {\it measured} event rate, corrected for the detection efficiency, is
\begin{equation}
\Gamma = \frac{1}{N_{obs}\Delta T_{obs}}\times\sum_{events}\frac{1}{\epsilon(t_{\mathrm{E}})}.
\end{equation}
This event rate and the duration distributions are used to constrain
the mass function and the kinematics of the lensing structures 
\cite{Griest91} 
\cite{Zhao98}. 
Such studies 
will only be briefly
mentioned in this review, as the available efficiency controlled statistics
are still very limited for quantitative comparisons.
\section{Observations}
Three types of programs can be distinguished.
The two first ones, that can be called ``survey programs'', monitor
large samples of stars that are either identified in the images
(catalog survey programs) or that are not identified
(pixel survey programs). The third type can be considered as ``follow-up
programs'' that finely monitor individual events detected by the early warning
systems of the survey programs.
In this review, we will only discuss the catalog survey programs
with brief mentions of the other types that are discussed in details
in the other contributions of this volume;
in particular, results from the M31 pixel surveys are discussed in the Calchi
Novati's review \cite{M31-Novati}, and the follow-up
programs are reviewed by M. Dominik \cite{Dominik2009}.

Table \ref{tab:setups} lists the characteristics of the devices
used in the catalog survey programs.
\begin{table}
\begin{center}
\caption[]{Experimental setups devoted to microlensing surveys,
with the telescope diameter, the list of the filters, the total number of pixels,
the field covered by the detector, the list of surveyed targets
with the corresponding monitored fields, numbers of monitored stars,
approximative average sampling and observation epochs.
The indicative sampling values are averaged over the observation
seasons or during the observation nights.

Note: the MACHO SMC data were lost during the Mount Stromlo fire
of 18 January 2003.
}
\label{tab:setups}
\begin{tabular}{@{}c|c|c|c|c|c|c|c|c@{}}
\hline
          &telesc. 	& pixels       	& detector          &        & field    & stars       & approx.  & obs.\\
   Set-up & filters  	& $\times 10^6$ & field ($\degree$) & target & $deg^2$ 	& $\times 10^6$ & sampling & dates\\
\hline
{\bf EROS1} & 1 m & 1000 & $5.2\!\times\! 5.2$ 		& LMC & $27.$ &$4.2$ & 3 days & 90-93\\
{\bf plates} & R/B & & & & & & & \\
\hline
{\bf EROS1} & 0.4 m & 4 & $0.4 \!\times\! 1.1$ 		& LMC & $0.44$ & $0.1$ & 20 min. & 91-94\\
{\bf CCD}   & R/B  &   &                  		& SMC & $0.44$ & $0.1$ & 20 min. & 93-95\\
\hline
{\bf EROS2} &  1 m & $2\! \times\! 32$ & $0.7\! \times\! 1.4$ & LMC & $84.$ & $29.2$ & 3 days & 96-02\\
	   & I/V  & 		    &                 	& SMC & $9.$ & $4.2$ & 3 days & 96-02\\
	   &      & 		    &                 	& GC & $66.$ & $60.$ & 3 days & 96-02\\
	   &      & 		    &                 	& GSA & $20.1$ & $12.9$ & 3 days & 96-02\\
\hline
{\bf MACHO} &  1.27 m & $2\! \times\! 16$ & $0.7\! \times\! 0.7$	& LMC & $14.7$ & 11.9 & 4 days & 92-99\\ 
           &  R/B      &    	   & 		     		& SMC & $3.$   & 2.2 & 3 days & 93-96  \\
           &         &    	   & 		     		& GC  & $48.$  & 50.2  & 3 days & 92-99 \\
\hline
{\bf OGLE I} &  1 m & 4  & $0.25\! \times\! 0.25$  & GC & $1.25$ & 1.5 & 3 days & 92-95\\
           &   I  & & & & & & & \\
\hline
{\bf OGLE II} &  1.3 m & 4 & $0.24\! \times\! 0.95$ & GC & $11.$ & 20.5 & 3 days & 97-00\\
            &  I/B/V &   & drift-scan         & LMC & $4.7$ & 5.5 & 3 days & 96-00\\
            &        &   & 	             & SMC & $2.4$ & 2.2 &  & \\
\hline
{\bf OGLE III} &  1.3 m & 64 & $0.58\! \times\! 0.58$ & LMC & $38.$ & total 
& 2 days & 01-09\\
           &   I/V &    &                & SMC & $13.$ & 120.  & & \\
           &        &    &                & GC & $35.$ &     & & \\
\hline
{\bf DUO} &  1 m & 1000 & $5.2\! \times\! 5.2$ & GC & $27.$ & 13. & 1 day & 94 \\
{\bf plates} &  R/B   & & & & & &  & \\
\hline
{\bf MOA1-} 	&  0.61 m & 9 & $(1.\! \times\! 1.)/4$ & LMC  & 3.5 & 1.  & 3 hours & 95-98 \\
{\bf cam1}	&  B/R	& 	& & SMC  & 2.8 & 0.4 & 3 hours & 95-98 \\
\hline
{\bf MOA1-} &  0.61 m & 24 & $0.92\! \times\! 1.38$ & GC  & $20.$ & 5. & 2 hrs & 98-05 \\
{\bf cam2} &  B/R    &    &              	    & LMC & $20.$ & 3. & 1 day & 98-05\\
            &         &    &              	    & SMC & $10.$ & 1. & 1 day & 98-05\\
\hline
{\bf MOA2-} &  1.8 m & 80 & $1.32\! \times\! 1.65$ & GC & $48.$ & 50. & 1 hr & 05-\\ 
{\bf cam3} & R/V/I  &	 &			  & LMC & $31.$	& 50. & 1 hr & 05-\\
           &   	    &	 &			  & SMC & $4.4$	& 5.  & 1 day & 05-\\
\hline
{\bf Super-} & 4 m & 64 & $0.57\! \times\! 0.57$ & LMC & $22.$ & 100. & 2 days & 01-05 \\
{\bf MACHO} & VR   &    & 		       &  & & & & \\
\hline
\end{tabular}
\end{center}
\end{table}
%
The common requirements of these surveys are the monitoring of
a very large number of stars, with the best time-sampling
and the best photometric resolution that can be achieved.
These requirements drive the main specifications of the instruments:
they all monitor wide crowded fields
with broad passband filters --to get a large photon flux.
The achromaticity of the microlensing events further supports this
use of wide passband filters; since the ratio of luminosities does
not depend on the wavelength, the signal significance increases
with the photon flux. These surveys use in general at least two different
passbands to make easier the identification of achromatic events.
Figures \ref{fieldsGal} and \ref{fieldsLMC} show the fields monitored by the
different teams towards the Galactic plane and the Large Magellanic Cloud.
\vspace{-2mm}
\paragraph{\bf EROS: Exp\'erience de Recherche d'Objets Sombres.}
After a first phase started in 1990, using ESO-Schmidt
(1m) telescope photographic plates digitized with the MAMA machine
and a 16 CCDs camera \cite{erpre} \cite{ccdcam},
the EROS team used the 1m (F/5) MARLY telescope
installed at La Silla observatory (Chile) from July 1996 to February 2003,
with a dichroic beam-splitter allowing the simultaneous imaging in
two wide pass-bands (EROS-blue and EROS-red)
of a $0.7 \degree (\alpha) \times 1.4 \degree (\delta)$
field \cite{Bauer}.
Photons were collected by two cameras equipped with eight
$2K\times 2K$ LORAL CCDs each. 
Exposure times ranged typically from 2 to 5 minutes
and were optimized to
maximize the global significance of the photometric measurements
taken during microlensing magnifications.
EROS monitored LMC and SMC, the Galactic Center (GC), and
four directions in the Galactic Spiral Arms (GSA).
\vspace{-2mm}
\paragraph{\bf MACHO: MAssive Compact Halo Objects.}
From June 1992 to January 2000 the MACHO team used a 1.27 m (F/3.9) dedicated
telescope \cite{machotelescope}, also equipped with a dichroic
beam-splitter and two cameras of four $2K\times 2K$
LORAL CCDs each \cite{machocamera1} \cite{machocamera2}.
This setup, installed at Mount Stromlo (Australia),
provided simultaneous blue and red band photometry
and covered a field of
$0.7 \degree (\alpha) \times 0.7 \degree (\delta)$
with a pixel size of 0.62 arcsec.
The LMC, SMC as well as the Galactic
Center have been monitored for microlensing searches.
The observing strategy changed during the project and
the sampling of the Galactic fields varied from a few
observations per year to daily observations.
\vspace{-2mm}
\paragraph{\bf OGLE: Optical Gravitational Lensing Experiment.}
After a first phase started in 1992 \cite{OGLE1},
making use of a 1m (F/7)
telescope in Las Campanas (Chile) to monitor the Galactic
Center, 
the OGLE team has built its
own dedicated instrument, using a 1.3m (F/9.2) telescope,
equipped with a
thinned CCD camera of $2K\times 2K$ pixels and a filter wheel
with standard UBVRI filters \cite{ogle2tel}.
The system operated
in drift-scan mode with the telescope drifting in declination
at a rate of a few arc-seconds per time second.
In these conditions a single image covered a field of
$0.25 \degree (\alpha) \times 0.9 \degree (\delta)$
with a pixel of 0.4 arcsec.
LMC, SMC and the Galactic bulge have been monitored with this new setup
from 1997 to 2000.
Each field was observed every third night in I-band by OGLE-II,
and every 11-th night in the V-band.
OGLE-III had a $8K\times 8K$ pixel camera, made of 8 CCDs.
Observations were performed towards SMC, 
LMC and the Galactic bulge \cite{OGLE3-realtime} from
June 2001 to May 2009.
OGLE-IV is made of 32 E2V CCDs with $2K\times 4K$ pixels each
and received its first light in September 2009.
\vspace{-2mm}
\paragraph{\bf DUO: Dark Unseen Objects.}
This search for microlensing towards the Galactic Center, also using
the ESO-Schmidt telescope photographic plates digitized with
the MAMA, was performed in 1994 \cite{DUO1}.
\vspace{-2mm}
\paragraph{\bf MOA: Microlensing Observations in Astronomy.}
This team started microlensing searches 
with a 0.6m telescope located at mount John Observatory (New-Zealand),
equipped with an array of $3\times 3$ non-buttable $1K\times 1K$ CCDs \cite{MOAcam1}. Two different focal lengths were used during this
first stage.
The second camera had three $2K\times 4K$ pixels thinned SITe CCDs \cite{MOAcam2}.
The team is now using a new 1.8m telescope, with ten $2K\times 4K$ pixels E2V CCDs \cite{MOAcam3}
and it performs a real-time difference imaging analysis as described
in \cite{MOA-real-time}.
\vspace{-2mm}
\paragraph{\bf SuperMACHO.}
From October 2001 to 2005, this collaboration used
the 4m CTIO Blanco telescope during 150 half-nights,
equipped with eight SITe
$2K\times 4K$ CCDs and a $250nm$ passband VR filter,
to monitor all star types in the LMC \cite{supermacho1}.
Considering the serious supernov{\ae} contamination, complementary
observations triggered through an alert system were
used to identify microlensing events.

\section{Analysis techniques}
The data analysis of the surveys are all based on the same principles :
Firstly, light curves ({\it i.e.} flux versus time curves)
for cataloged stars or for emerging pixel clusters are
extracted from the raw data, using automatic astrometric
and photometric alignment procedures; secondly,
each light curve is subjected to a microlensing
search algorithm based on the expected characteristics
of the events.
\subsection{Photometry}
Simple aperture photometry is not adapted for crowded fields.
The microlensing pioneers have used or developed Point Spread Function
(PSF) photometry techniques to extract the luminosity of each
star in crowded fields (\cite{Eros1plaqlim} and references therein
for the photographic plates, \cite{DoPhot} \cite{PEIDA} for the CCD).
A simultaneous fit of the PSF is performed on stars that are
not completely separated.
The following generation of microlensing teams have developed
Difference Image Analysis techniques (DIA) \cite{Alard} \cite{Wozniak} \cite{leguillou} \cite{MOA-real-time}
that are very demanding in computing ressources, and are now usable
thanks to the advances in computers.
The base of this technique is the construction of the image difference
between a current and a reference images. A convolution kernel is
calculated and used to convolve the reference image to match the
seeing of the current image. The two images are then subtracted and
the variability are searched from the study of the residuals.
This technique has the advantage to make detectable the magnification
of objects that are not detected when unmagnified.
\subsection{Production of light-curves}
Catalogs of objects are extracted from
reference images, produced by the co-addition of series of the
best quality images.
For PSF photometry, the light-curves are produced after geometric
and photometric alignment of current images on reference images.
For DIA technique, a list of variable objects is extracted from
the subtracted images and subsequently monitored in the other
images.
\subsection{Selection of microlensing candidates}
The general philosophy for this discriminant analysis is as follows:
if the starting point is a series of light-curves, the first
stage consists in selecting the less stable objects, then to search for
microlensing candidates through the topological characteristics
of the expected light-curve. 
All analysis include a cleaning procedure, where
bad quality images (taken under poor atmospheric conditions, with guiding
problems...) are removed, sources with problematic environment
(such as dead pixels or bright neighboring stars) are discarded from the reference catalog,
and the identified aberrant measurements are suppressed.
Various algorithms of discriminant analysis
have been developed to pre-select light-curves with single positive
fluctuations, visible in each color (in the case
of multi-wavelength observations).
Microlensing fits are performed on the pre-selected
light-curves and the final selection is usually based on the
fitted parameters.

The last selection stage is the background rejection and
the residual background estimate.
This stage has long suffered from being somewhat underestimated.
Apart from the known experimental artifacts such as contamination
by egrets or those induced by the SN1987A echoes \cite{echoSN}, a
few categories of objects that could mimic a microlensing
event have been identified by the search teams
over the years;
the microlensing science would certainly benefit from a better understanding
of these fakes:
%

- The so-called ``blue bumpers'' \cite{MACHO2yrs}, bright blue
variable stars that show recurrent bumps. Amongst them, Be type stars
that are known to be unstable (case of the EROS1-1 candidate \cite{Beaulieu})

- Supernov{\ae} that explode in galaxies behind the monitored target \cite{macho2000LMC}.
%
\subsection{The efficiency control}
\label{sec:efficiency}
The optical depth calculations require the knowledge
of the detection efficiency as a function
of $t_{\mathrm{E}}$, averaged over the impact parameter space,
over the observation period and over the
source apparent magnitudes and colors.
This detection efficiency is defined as the ratio of
the microlensing events that satisfy the selection requirements
to the number of events with $u_0<1$
whose magnification reaches its maximum during the observing period.
Its calculation needs the simulation of events in a domain of the
parameter space that exceeds by a large amount
the expected domain of sensitivity
(in EROS, $u_0$ was generated up to 2, $1\ day<t_{\mathrm{E}}<900\ days$, $t_0$ was
generated from 150 days before the first observation to 150 days
after the last one).
The simplest technique to simulate microlensing-like light-curves
consists in superimposing simulated events on
measured light curves from an unbiased sub-sample of the monitored catalog.
Events are simulated as point-source, point-lens, constant velocity
microlensing events.
This technique automatically takes into account the
real time sequence,
the source variabilities and the photometric systematic or
accidental distortions that are preserved when combining with
a simulated microlensing through rescaling procedures.
The simulation of the microlensing in series of synthetic images (instead
of light-curves) have also been made by \cite{macho2000bulge} and \cite{Sumi2006},
followed by the full photometric reduction and light-curve production.
MOA used an intermediate technique of superimposing a microlensing
effect on real subtracted images \cite{MOA-tau-CG}.

When the monitored population has a large distance dispersion, then
the microlensing detection efficiency for a given
stellar type depends on
the distance, since it depends on the apparent magnitude.
This makes the efficiency to be
ultimately correlated with the optical depth itself.
As the microlensing detection efficiency essentially depends on the
{\it apparent} magnitude and possibly on the color,
the mean optical depth up to a given magnitude (or for a given
catalog) $\tau_{cat}$ is the concept that
has to be used in the case of microlensing of populations
with a wide distance distribution such as those belonging
to the Galactic Spiral Arms \cite{GSA7y}.
Such a measured optical depth $\tau_{cat}$ can be compared
with the predictions derived from a lens and source
distribution model using expression (\ref{optpop}),
where the observed source distance distribution $\ud n_S(D_S)/\ud D_S$
corresponds to the cataloged sources (taking into account
the star detection efficiency).
\subsubsection{Blending}
\begin{figure}
\begin{center}
\includegraphics[width=5.cm]{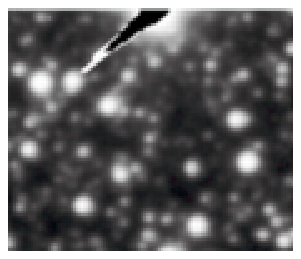}
\includegraphics[width=5.cm]{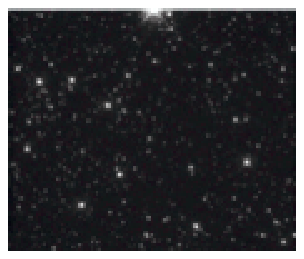}
\caption{\it
(left) The $R_{EROS}$ composite image (used to detect the cataloged stars) and
(right) the U6FQ1104B-HST image of the same sub-field towards gs201.
\label{blending}}
\end{center}
\end{figure}

Due to source confusion, single apparent {\it cataloged objects}
can be made of several {\it stars}.
This fact, called blending,
is one of the major sources of uncertainty for the optical depth
determination in crowded fields, as it affects not only the
microlensing detection
efficiency but also the effective number of monitored sources $N_{obs}$
that enters in expression (\ref{taumeas}) \cite{blending-Stefano} \cite{blending-Alard} \cite{blendingOGLE} \cite{blending-Han}.
When some blending is taking place,
the reconstructed stellar flux contains contributions from more than
one star. For this reason, the effective number of monitored sources $N_{obs}$
that should enter in expression (\ref{taumeas}) is larger than the
number of cataloged stars. Also, as only one of the components of the blend
is magnified, this situation makes the reconstructed
apparent magnification of the blend smaller than the real magnification
of the lensed component. As a consequence, the detection efficiency
appears smaller than for the unblended case.
The impact of blending on the optical depth estimates
has been studied by the observational teams using different approaches;
as an example \cite{afonso2003} \cite{Hamadache} did a study
based on synthetic images and on real images enriched with
randomly located synthetic stars.
In this type of simulation, it is usually
assumed that there is no spatial correlation between the blend components
and that the blend results only from a random coincidence.
As the existence of correlation should increase the blending,
this should be systematically checked and taken into account if
necessary for the simulations.
Other studies \cite{macho2000LMC} \cite{ogletauLMC} \cite{GSA7y}
compare HST images with the corresponding monitored subfields.
For all these studies, the critical ingredient is
the underlying luminosity function.
Fig. \ref{blending} shows better than in words what is
the blending in a Galactic plane field, and illustrates the
relation between the (HST) luminosity function and the
content of an EROS object.
The HST images that are deeper by at least 2-3 magnitudes
than the ordinary survey catalogs and have a PSF of $\sim 0.1''$
---10 times smaller than the terrestrial images---
allow one to extract relevant statistical data
in the underlying ingredients of the objects cataloged by the surveys.

A correction to the efficiency estimates is derived from these studies.
There is some compensation between the loss
of event-by-event efficiency and the gain in the effective number of monitored
objects, that limits the impact on the optical depth estimates.
Finally, the systematic
uncertainty on the optical depth due to blending is generally
estimated to be smaller than $10\%$. Decreasing this uncertainty needs
a careful estimate of the blending corrections.
The most simple way to minimize the impact of blending is to keep
only microlensing of red giant (bright) stars.
This option has been used by all the search teams towards the Galactic Center,
and by EROS and OGLE towards the LMC.
\subsubsection{The efficiency uncertainties due to complex events}
The efficiency estimates are based on the simulation of simple 
microlensing events. It has been estimated
that about $10\%$ of the real microlensing events
have a more complex 
evolution such as a double lens configuration or a parallax
effect \cite{Glicens}.
In the later case, the correction is usually small, but in the first
(rare) case, the mean detection efficiency may be seriously affected by
the distorted shape of the light-curve \cite{Glicens}.
The impact of this detection modification is sometimes
accounted for as an extra-uncertainty of $\sim 10\%$
on the detection efficiency.
\subsection{Complementary observations and statistical tests}
\label{sec:complementary}
These {\it a posteriori} tests are crucial not only to
check and improve the background rejection, but also to gain
confidence in the statistical distributions.
The first microlensing candidates have been investigated in
details through complementary photometric and spectroscopic
measurements. A non negligible fraction of the early candidates
(with low signal to noise) has been rejected after publication
because of the late discovery of anomalies such
as repetition of bumps ({\it e.g.} the so-called
``blue-bumpers'' \cite{MACHO2yrs})
or baseline variability \cite{Eros1cand2}.
In connection with optical depth studies, systematic
statistical tests are performed on the populations of
detected microlensing candidates to check that the impact
parameter $u_0$ and the maximum magnification time $t_0$
distributions ---with uniform prior--- are only
biased by the detection efficiencies, and that this
population is representative of the monitored population
after correcting for the detection efficiencies.
\section{Some limitations and difficulties in the optical
depth determination and future improvements}
\subsection{Statistical limitations}
The impact of poor statistics is worsen by the
dispersion of the microlensing event durations;
since the variability of the durations induces a strong variability of
the individual contributions to the optical depth,
the statistical uncertainties on this parameter
are indeed larger than expected from Poisson
statistics. This is taken into account by the search teams
by using the procedure described in \cite{Han-1995}.

One of the statistical limitations in the microlensing
surveys is the available number of sources that can be monitored.
As an example, optical depth measurements towards
obscured regions of the Galactic plane are very strongly
limited by the visibility of the sources. The only way to
improve the situation is to make infrared surveys \cite{Gould95}.
The VVV (VISTA Variables in the Via Lactea) program \cite{VVVproposal},
one of the VISTA\footnote{Visible and Infrared Survey Telescope for Astronomy} 
\cite{VISTA} large survey programs, aims at searching for
variabilities towards the Galactic plane and should
have a major impact for the microlensing searches
towards the dusty regions.

Another statistical limitation ---more difficult to overcome---
comes from the weakness of the signal itself;
more exposure ($N_{obs}\times \Delta T_{obs}$) is needed to compensate for
the low microlensing rate towards the Magellanic Clouds.
SuperMACHO \cite{supermacho1} which monitored the largest possible number of stars
within the LMC was specifically devoted to this search.
\subsection{Information from simple events}
The ordinary microlensing events
provide little physical information on the lens configuration.
Any technique improving the photometric accuracy through
the use of larger telescopes or by achieving high sampling rates
increases the chance to get some extra-information by the
detection of small deviations with respect to the simple
microlensing light-curve.
International networks of relatively small telescopes
linked to the alert systems of surveys
like PLANET\footnote{Probing Lensing Anomalies NETwork} 
\cite{webPLANET} MicroFUN\footnote{Microlensing Follow-Up Network} 
\cite{webMicroFUN} or MINDSTEp\footnote{Microlensing Network for
the Detection of Small Terrestrial Exoplanets} \cite{webMINDSTEp}
have developed the second approach.

The LSST\footnote{Large Synoptic Survey Telescope}
project \cite{LSSTscbook} will certainly dramatically change the
landscape for the photometric accuracy, as the telescope
has an equivalent diameter of 6.5m.
Tens of thousands of events should be detected each year
with a sampling rate of once every few nights.
Considering the photometric accuracy, parallax and extended
source effects will often be detectable and improve the
knowledge of the lens configuration;
but high sampling networks will be needed in the case of
very fast variations during the caustic crossing of double lens events.
\subsection{Limitation from the knowledge of the source distance distribution}
The Magellanic Clouds are stellar populations located at
well defined
distances. Therefore, the allowed
parameter space of an observed lens configuration is
less extended than when the source distance is poorly known.
The distance distribution of the Galactic bulge population is 
wider, but the relative uncertainty on the bright source's
positions is less than $10\%$ \cite{distCG}. This uncertainty does not have
a strong impact on the optical depth estimates.
In contrast, monitored sources at large Galactic longitude can
span a wide range of distances (see Sect. \ref{sec:population}).
The interpretation of the mean optical depths requires either a
good knowledge of the source distance distribution (through complementary
observations such as spectroscopy) or a model that allows the
simulation of the survey catalogs and to predict the mean optical depths \cite{GSA7y}.
\section{Review of results}
\subsection{Galactic model}
\label{sec:model}
We describe below the model and the parameters of one of
the simplest Galactic models, that is commonly used for
the optical depth interpretations. To facilitate the
comparisons, we will interpret all the optical depth results
and some event duration distributions
within this framework, and briefly mention the alternatives
in the subsequent discussion.
The model includes a central bulge, a thin disk, a hypothetical thick disk
and a halo; since the exact morphology of the Galactic Spiral Arms (GSA) is 
yet unknown \cite{englmaier}, no spiral arm features have been included 
in the model.
The parameters we use here are summarized in Table \ref{tab:Gmodel}.
\begin{itemize}
\item
{\bf The observer position and velocity:}
To calculate the optical depths and duration distributions within the
Galactic plane, one needs to take into account the position of the
Sun with respect to the disk ($15.5\pm 3\, pc$ above the median plane,
according to \cite{positionsun})
and the peculiar solar motion with respect to the Local Standard
of Rest \cite{propermotion}:
\begin{equation}
v_{\odot R}=10.0,\ v_{\odot \theta}=5.3,\ v_{\odot z}=7.2\ \
({\rm km}{\rm s^{-1}}).
\end{equation}
\item
{\bf The Galactic halo:}
Many models of the Galactic halo have been tested to interpret the
observations; from the simple ``standard'' spherical halo to
flattened halo. Discussions can be found in \cite{Binneyetal}
\cite{Dwek} \cite{freudenreich}, and the corresponding predicted
optical depth maps can be found in \cite{evans}.
We will only consider here the so-called S-model \cite{MACHO2yrs}
that allows the
easiest comparisons and combinations between the various published
results towards LMC and SMC.
This spherical halo is considered as isotropic and
isothermal with a density distribution given in
spherical coordinates by~:
\vspace{-2mm}
\begin{equation}
\rho_{H}(r) = \rho_{h\odot} \frac{R_{\odot}^{2}+R_{c}^{2}}{r^{2}+R_{c}^{2}}\ ,
\end{equation}
where $\rho_{h\odot}$ is the local halo density, $R_{\odot}$ is the Galactocentric
radius of the sun,
$R_{c}$ is the halo ``core radius'' and $r$ is the Galactocentric radius.
A Maxwellian distribution is assumed for the velocities with a one-dimensional
dispersion $\sigma_{1D}$.
Usually authors
consider mono-mass distributions for the lenses when simulating microlensing.
\begin{table}
\begin{center}
\caption[]{Parameters of the Galactic model. 
\label{tab:Gmodel}
}
\begin{tabular}{|c|l|c|c|}
\cline{2-4}
\multicolumn{1}{c|}{}	& $R_{\odot}\ ({\rm kpc})$	& \multicolumn{2}{c|}{8.5} \\
\hline
	&					& Thin	& Thick	\\
\cline{3-4}
        & $\Sigma_{\odot}\ (M_{\odot} {\rm pc}^{-2})$   & 50. & 35. \\ 
        & $H\ ({\rm kpc})$                      & 0.325 & 1.0 \\ 
       	& $h\ ({\rm kpc})$                      & 3.5 & 3.5\\
Disks        & $M_{thin}(\times 10^{10}M_{\odot})$   & 4.3 & 3.1 \\
        & $\sigma_r\ (km\ s^{-1})$              & 34. & 51. \\
        & $\sigma_{\theta}\ (km\ s^{-1})$       & 28. & 38. \\
        & $\sigma_z\ (km\ s^{-1})$              & 20. & 35. \\
\hline
        & $a\ ({\rm kpc)}$      		& \multicolumn{2}{c|}{1.49}     \\ 
        & $b\ ({\rm kpc})$      		& \multicolumn{2}{c|}{0.58}     \\ 
Bulge   & $c\ ({\rm kpc})$			& \multicolumn{2}{c|}{0.40}       \\
	& Inclination $\Phi$			& \multicolumn{2}{c|}{$45 \degree$}      \\
        & $M_{B} (\times 10^{10}M_{\odot})$     & \multicolumn{2}{c|}{1.7} \\
        & $\sigma_{bulge}\ (km\ s^{-1})$        & \multicolumn{2}{c|}{110.} \\ \hline
	& $\rho_{h\odot}\ (M_{\odot} {\rm pc}^{-3})$ & \multicolumn{2}{c|}{0.0078}  \\
Halo	& $R_{c}\ ({\rm kpc})$                	& \multicolumn{2}{c|}{5.0}  \\ 
S-model	& $\sigma_{1D}\ (\rm km.s^{-1})$	& \multicolumn{2}{c|}{155.} \\
	& $M\ in\ 60 \ {\rm kpc}\ (10^{10} M_{\odot})$   & \multicolumn{2}{c|}{40} \\ \hline
\hline
	& $\rho_{\odot}\ (M_{\odot} {\rm pc}^{-3})$	& 0.085	& 0.098 \\
Predictions	& $V_{rot}\ at\ sun\ ({\rm km}\ {\rm s}^{-1})$   & 211 & 222  \\ 
	& $\tau_{LMC}$	& $4.7\times 10^{-7}$ &	\\      
	& $\tau_{SMC}$	& $6.58\times 10^{-7}$ &	\\ \hline    
\noalign{\smallskip}
\end{tabular}
\end{center}
\end{table}
\item
{\bf The Galactic disks:}
The disk densities are modeled by a double exponential expressed
in cylindrical coordinates:
\begin{equation}
\rho_{D}(R,z) = \frac{\Sigma_{\odot}}{2H} \exp 
\left(\frac{-(R-R_{\odot})}{h} \right) \exp 
\left( \frac{-|z|}{H} \right) \ ,
\end{equation}
where $\Sigma_{\odot}$ is the column density of the disk at the Sun position, 
$H$ the disk height scale and $h$ its length scale.
%
A thick disk has been discussed as a possible
component of the Galactic structure 
\cite{Grenacher-1999} \cite{Reid} 
\cite{GSA2y} \cite{GSA3y} \cite{GSA7y},
for the interpretation of the optical depth towards LMC/SMC and
far from the Galactic center.
The kinematics of these disks involves a global rotation
given by \cite{rotdisc}
\vspace{-2mm}
\begin{equation}
V_{rot}(r) = V_{rot,\odot} \times \left[ 
1.00762 \left( \frac{r}{R_{\odot}} \right)^{0.0394} + 0.00712 \right] \ ,
\end{equation}
where $V_{rot,\odot} = 220$ {\rm km/s} \cite{rotdisc},
and an anisotropic Gaussian peculiar velocity distribution,
characterized by the velocity dispersions of Table \ref{tab:Gmodel}.
The mass function for the lenses is taken
from \cite{Gould-1997} for both the thin disk and the bulge.
\item
{\bf The Galactic bulge:}
The density distribution for the bulge -~a bar-like triaxial model~-
is taken from \cite{Dwek} model G2, given in Cartesian 
coordinates:
\begin{equation}
\rho_{B} = \frac{M_{B}}{6.57 \pi abc} e^{-r^{2}/2} \ , \ 
r^{4} = \left[ \left( \frac{x}{a} \right)^{2} + 
	       \left( \frac{y}{b} \right)^{2} \right]^{2} + 
	\frac{z^{4}}{c^{4}} \ ,
\end{equation}
where $M_{B}$ is the bulge mass, and $a$, $b$, $c$ the length
scale factors.
The inclination of the bar 
\cite{orientationbar} differs with the authors
(from $\sim 14\degree$ according to 
\cite{freudenreich} to $45\degree$ according to 
\cite{Hamadache} and \cite{picaud}).
The transverse velocity distribution of the bulge stars is given by
\begin{equation}
f_{T}(v_{T}) = 
\frac{1}{\sigma_{bulge}^{2}} v_{T} \exp \left( -\frac{v_{T}^{2}}{2\sigma_{bulge}^{2}} 
\right).
\end{equation}
\end{itemize}
\subsection{Results towards the Galactic plane}
\label{sec:population}
\begin{figure}
\begin{center}
\includegraphics[width=11.5cm]{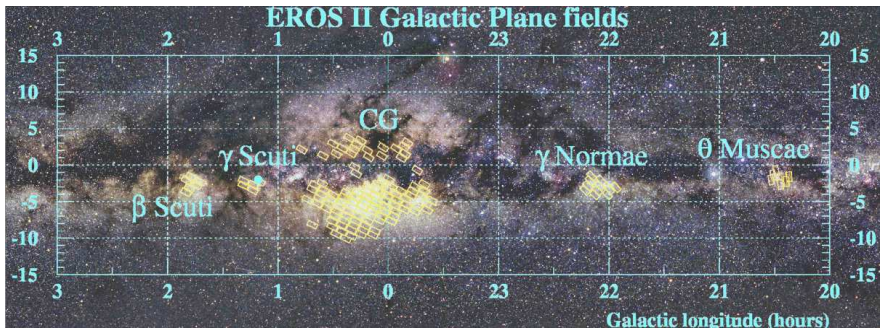}
\caption[]{\it View of the Galactic plane and fields monitored by EROS.}
\label{fieldsGal}
\end{center}
\end{figure}

\begin{figure}
\begin{center}
\includegraphics[width=10cm]{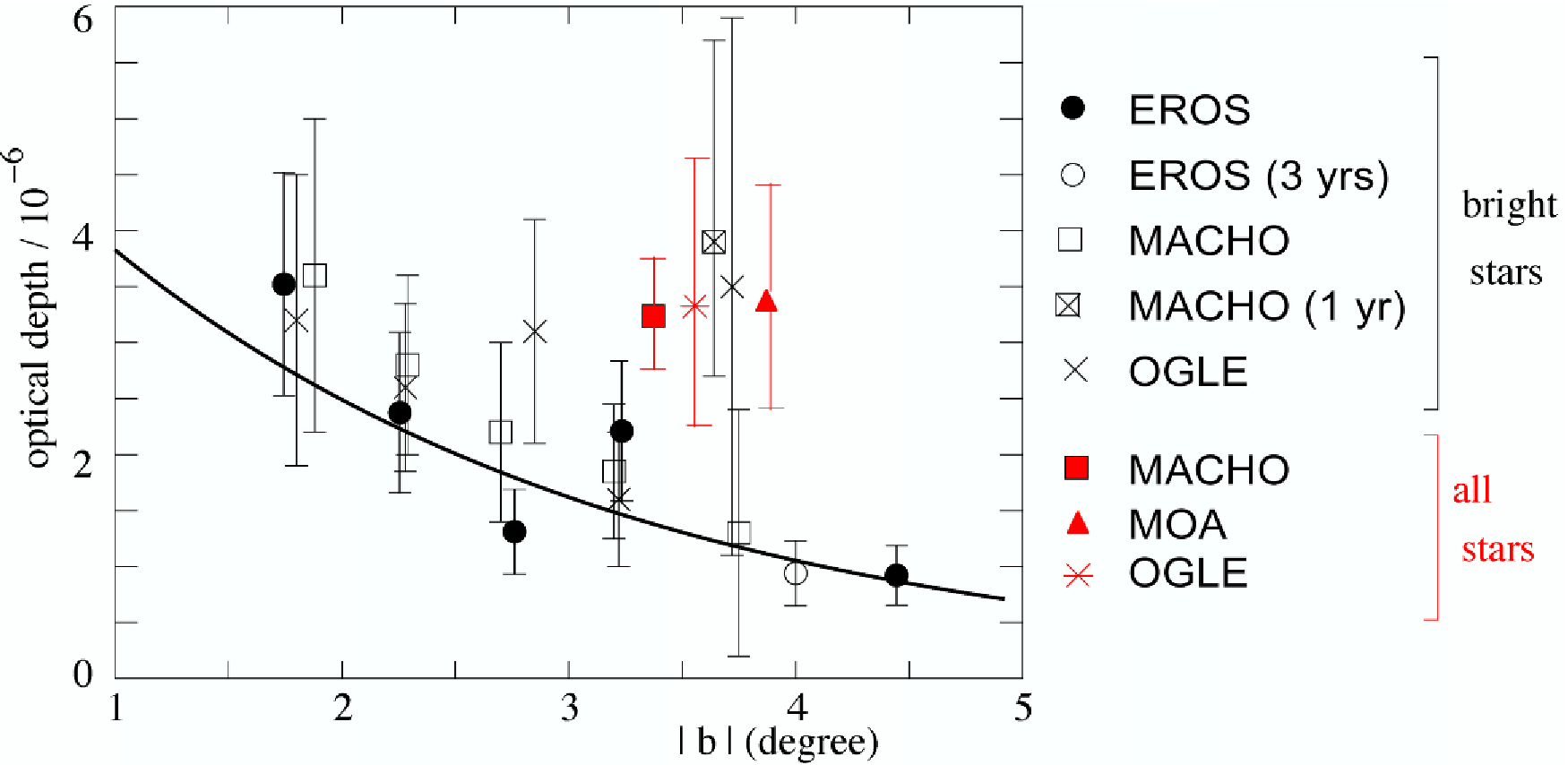}
\includegraphics[width=12cm, height=6cm]{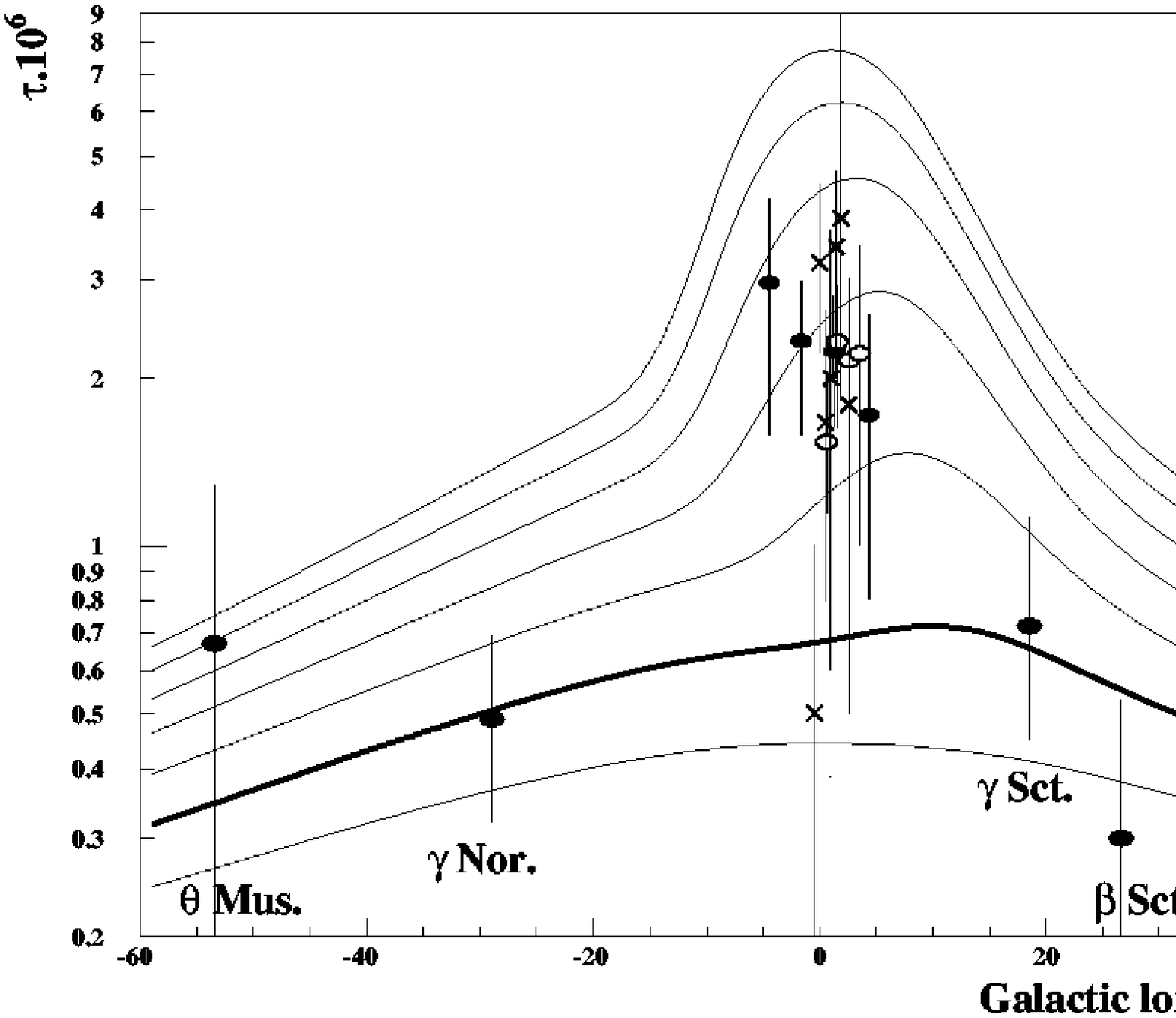}
\caption{\it
(up) The measured optical depth in the Galactic plane as a function
of the absolute Galactic latitude $|b|$.
The line shows the fit (\ref{profoptfit}).
(down) The optical depth as a function of Galactic longitude.
Results at large longitude $|l|>8\degree$ come from the
specific EROS GSA search at $b\sim -2.5\degree$.
Results around longitude zero (MACHO: open circles, EROS: filled
circles, OGLE: crosses)
come from the Galactic center searches at $|b|\sim 2.5\degree$.
The lines show the predicted optical depths as a function of
latitude  at 6,7 (thick line), 8, 9, 10, 11 and 12 kpc
(from lowest to highest curve) at $b=-2.5\degree$,
according to the model described in Sect. \ref{sec:model}.
The measured optical depth around $l=0\degree$ are compatible with the
expected optical depth at $8.5\ kpc$; the best
estimated mean distance for the sources in the Galactic Spiral Arms is
confirmed to be $7\, kpc$.
\label{resultatsG}
}
\end{center}
\end{figure}

Figure \ref{fieldsGal} shows the fields monitored by EROS
towards the Galactic plane. The fields monitored by MACHO,
OGLE and MOA are all included within the bright zones
of the Galactic center. MACHO also monitored fields close to
$\gamma Scuti$ ($l\sim 18\degree$) and OGLE monitored fields at
$l\sim -20\degree$ and $l\sim -30\degree$, but only EROS published
optical depth measurements far from the Galactic Center ($|l|>8\degree$).
More than 4000 events have been detected towards the Galactic bulge 
(but only a fraction of them have been used to obtain optical depths
under controlled efficiency),
and 27 towards the Galactic Spiral Arms (22 used for optical depth determination).
The interpretation of the optical depth measurements
is somewhat complicated by the limited knowledge of
the lensed source distances.
Indeed, the optical depth estimated for the full sample of
detected stars or with DIA analysis is the average
over all the stars across the lines of sight of the GC. The
optical depths towards the spiral arms also correspond to
average values.
The uncertainty on the source distances is reduced towards the Galactic center
by selecting the clump giant stars, with the triple advantage
to monitor a well located population ($D_S=8.5\; kpc \pm 10\%$),
less affected by blending problems, and benefiting from 
a better photometric accuracy because of the brightness of the sources.
The sources monitored towards GSA span a wide range of distances
($\pm 5$ kpc) and their mean distance ---estimated to be
$7\pm 1\,kpc$--- is also somewhat uncertain \cite{GSA7y}.
In the latter case, the measured optical depth towards the monitored
population corresponds to expression (\ref{optpop})
where $\ud n_S(D_S)/\ud D_S$ is the distance distribution
of the {\it monitored} stars,
corresponding to the true distribution distorted by the
star detection efficiency.
Because of this limited knowledge of the source distances, the results are
compared with optical depth estimates at various distances.
The following optical depth measurements are summarized in Fig. \ref{resultatsG}
(here $<t_{\mathrm{E}}>$ is corrected for the detection efficiencies):
\begin{center}
\begin{scriptsize}
\begin{tabular}{r|c|c|c|c|c|c|c}
				& sea- & field    & stars	& events      &      & $<\tau>_{bulge}$ & $<t_{\mathrm{E}}>$ \\ 
reference			& sons & $deg.^2$ & analyzed 	& for $\tau$  & $\bar l\degree,\bar b\degree $ & $\times 10^{6}$ & corrected \\ 
\hline       
OGLE \cite{oglcg94}		& 2	& 0.81 & all		& 9  & $\pm 5,-3.5$ & $3.3\pm 1.2$ \\        
MACHO \cite{MachoBulge-1an}	& 1	& 12.  & all		& 45 & $2.55,3.64$ & $3.9^{+1.8}_{-1.2}$ & \\
MACHO \cite{macho2000bulge}	& 3	& 4.   & all/DIA 	& 99 & $2.68,-3.35$ & $3.23^{+0.52}_{-0.50}$ & \\
EROS \cite{afonso2003}		& 3	& 15.  & bright		& 16 & $2.5,-4.0$ & $0.94\pm 0.29$ & \\
MOA \cite{MOA-tau-CG}		& 1	& 18.  & all/DIA	& 28 & $4.2,-3.4$ & $3.36^{+1.11}_{-0.81}$ & \\ 
MACHO \cite{machobul2005}	& 7	& 4.5  & bright		& 62 & $1.5,-2.68$ & $2.17^{+0.47}_{-0.38}$ & $21.6\pm 3$ \\ 
OGLE \cite{Sumi2006}		& 4	& 5.   & bright		& 32 & $1.16,-2.75$ & $2.55^{+0.57}_{-0.46}$ & $28.1\pm 4.3$  \\
\cline{6-7}
EROS \cite{Hamadache}		& 7	& 66.  & bright		& 120 & \multicolumn{2}{c|}{GC map} & $28.3\pm 2.8$  \\
EROS \cite{GSA7y}		& 7	& 20.1 & all		& 22 & \multicolumn{2}{c|}{GSA map} & $48.\pm 9.$ \\ 
\end{tabular}
\end{scriptsize}
\end{center}
\vspace{1mm}

%
Many of these measurements are correlated because
of the large overlap between the survey's exposures
($N_{obs}\times \Delta T_{obs}$).
Merging these results would require a close collaboration
between the authors of these surveys.

The variation with the latitude deduced from the largest sample
(EROS) is well fitted by
\begin{equation}
\tau/10^{-6}=(1.62\,\pm 0.23)\exp[-a(|b|-3\degree)],
\; {\rm with}\; a=(0.43\,\pm 0.16)\,\deg^{-1}.
\label{profoptfit}
\end{equation}
This fit agrees with the latest
results of MACHO \cite{machobul2005} and OGLE-II \cite{Sumi2006}
that use only bright stars, and with the Galactic models of 
\cite{evans} and \cite{Bissantz}.
The estimates deduced from
DIA analysis as well as from searches using all the
detected stars correspond to average optical depth along the line
of sight of the GC. For a relevant comparison with the optical depth
involving bright stars only that are all supposed to be close to the
Galactic center, we plot in Fig. \ref{resultatsG} the optical depth $\tau_{bulge}$ corrected for
the parasitic disk-disk lensing. Nevertheless,
these ``all star measurements'' seem to be
somewhat systematically larger than the measurements restricted to
bright stars. The corrections
for disk-disk lensing may be a little too large, or the impact of disk bright stars
may have been underestimated.

A local excess of optical depth has been reported by \cite{machobul2005} with
a $2\sigma$ statistical significance,
but no conclusive evidence for clustering of events was found.
%
%
The measured $\tau_{GSA}$ agrees with the predictions of the simplest
Galactic model (Fig. \ref{resultatsG} down), providing no evidence
for a thick disk.

Detailed maps of the expected microlensing optical depth are now available 
\cite{Kerins} for the interpretation of the available and future observations
towards the Galactic bulge.
They provide optical depths up to a given magnitude.

The evaluation of the efficiencies needed by the optical depth determinations
allows us to correct any distribution in order to extract the prior distributions.
This has been used firstly to check the statistical properties of the events
as already mentioned in Sect. \ref{sec:complementary},
and then to obtain their prior $t_{\mathrm{E}}$ distributions.
These $t_{\mathrm{E}}$ distributions given by the three collaborations
have been recently analyzed by \cite{Calchi} to constrain the
slope of the Galactic Bulge mass function ($\sim m^{-\alpha}$) in the
main sequence (MS) range; the result from the MACHO data set
($\alpha_{MS}=1.7\pm 0.5$) was found compatible with the numbers
deduced from the OGLE and EROS data sets, and also
in agreement with previous estimates \cite{Zoccali}.

As a conclusion regarding the Galactic plane directions,
there is now a satisfactory agreement between the
optical depths and the efficiency corrected duration
distributions obtained by the various teams.
This agreement, as well as the agreement with the Galactic models,
give confidence in the reliability of the technique, which
is necessary to address the more controversial subject of the microlensing
towards the Magellanic Clouds.
\subsection{Results towards the Magellanic clouds}
\begin{figure}
\begin{center}
\includegraphics[width=9cm]{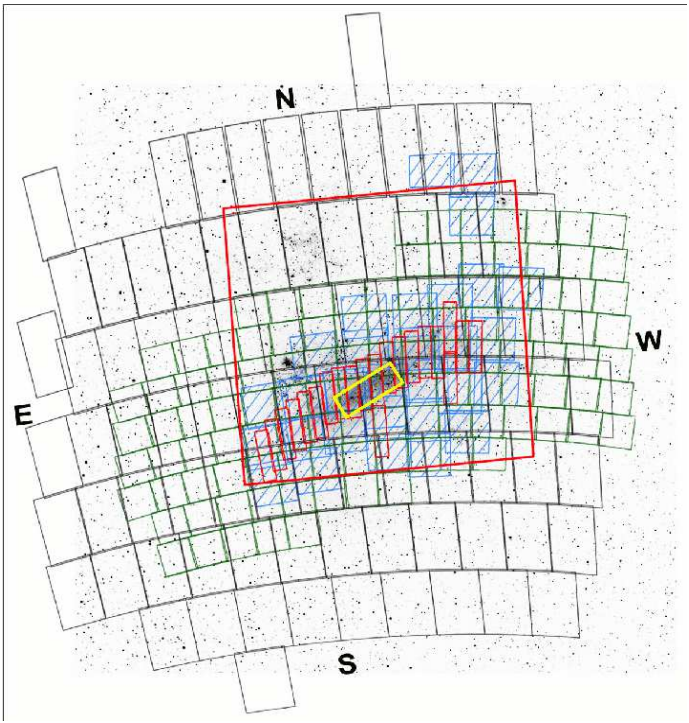}
\caption{\it
Approximate positions of the monitored LMC fields.
The red rectangles (resp. green squares) show the OGLE-II
(resp. OGLE-III) fields,
the blue dashed squares show the MACHO fields, the red
large square is the EROS1-plate field, the small yellow
rectangle within the LMC-bar indicates the EROS1-CCD field,
and the black rectangles show the EROS-2 fields.
\label{fieldsLMC}}
\end{center}
\end{figure}

Figure \ref{fieldsLMC} shows the fields monitored by the
different teams towards LMC.
The surveys of SMC are much more limited because of the size of this target.
The optical depths towards these directions have been studied by MACHO, EROS
and OGLE.
Detection of candidates has also been reported by the
MOA collaboration \cite{MOAcam1}, but without optical 
depth estimates.

\begin{figure}
\begin{center}
\includegraphics[width=9cm, height=5cm]{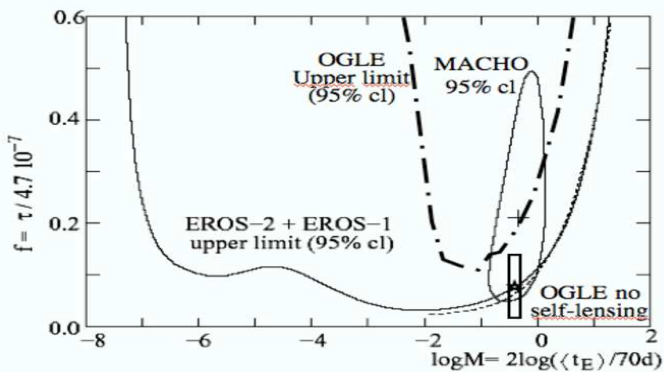}
\caption{\it
Published constraints on the fraction $f$ of the standard
spherical Galactic halo
made of massive compact objects as a function of their mass $M$.
The solid line labeled EROS shows the EROS upper
limit \cite{ErosLMCfinal}, based on the combination of the EROS1 results
with the LMC EROS2 results.
The dotted (incomplete) line takes into account the SMC search
direction, assuming that the observed candidate belongs to the halo.
The OGLE upper limit \cite{ogletauLMC} is shown in dashed-dotted line.
The closed domain is the $95\%$ CL contour for the
$f$ value claimed by MACHO and the star surrounded by the small rectangle
shows the
contribution of the 2 OGLE events, assuming that they belong to the halo.
\label{resultatsLMC}
}
\end{center}
\end{figure}

The main result from the LMC/SMC surveys is that
compact objects of mass within $[10^{-7},10]\times \Msol$
interval are not a major component of the
hidden Galactic mass (Fig. \ref{resultatsLMC}).

Light objects ($< 10^{-1}M_{\odot}$) in the Galactic halo were clearly ruled out as early as 1997
by EROS1 \cite{EROS1CCD1} which did not find any short time-scale events
(with $t_{\mathrm{E}}<1\ day$).
Heavy objects ($1$ to $30M_{\odot}$) contribution to the Galactic halo
was also seriously constrained as soon as 2001 by MACHO \cite{MACHOhighmass}.

The contribution of medium mass objects to the halo {\it has been} a
more controversial subject, as background sources as well as
self-lensing contributions \cite{Sahu-selflensing1} \cite{Sahu-selflensing2}
have complicated the data analysis and its interpretation.
The MACHO collaboration has published non-zero halo signal estimates towards
the densest part of the LMC, but EROS did not find indisputable events
within its sub-sample of bright stars in its wider LMC search field, and
has only published upper limits on the halo contribution.
To illustrate some of the difficulties of the LMC analyses, we have to
mention that three of the MACHO \cite{Bennett} and ten of the
EROS \cite{ErosLMCfinal} LMC/SMC early published candidates 
have been discarded as viable microlensing events because of a
significant loss of the signal to noise ratio from longer
light-curves or because of the detection of a second bump
(as late as 8 years later \cite{lasserre}).
%
\subsubsection{Exclusion limits on the Galactic halo}
Four results 
are mutually compatible:
\begin{itemize}
\item
the EROS upper limit \cite{ErosLMCfinal},
\item
the MACHO high-mass upper limit \cite{MACHOhighmass},
\item
the most recent OGLE-II upper limit \cite{ogletauLMC},
\item
and the first combination of data from different teams
\cite{EROSMACHO} (hereafter called EROS1-CCD+MACHO)
obtained with the EROS1-CCD high sampled data ($20$ minutes)
taken in 1991-95 with the 16 CCD camera
\cite{EROS1CCD1} \cite{EROS1CCD2} and with the MACHO-2year short
events search in 1992-94 through a spike analysis \cite{MACHOspike}.
\end{itemize}

\begin{figure}
\begin{center}
\includegraphics[width=10cm]{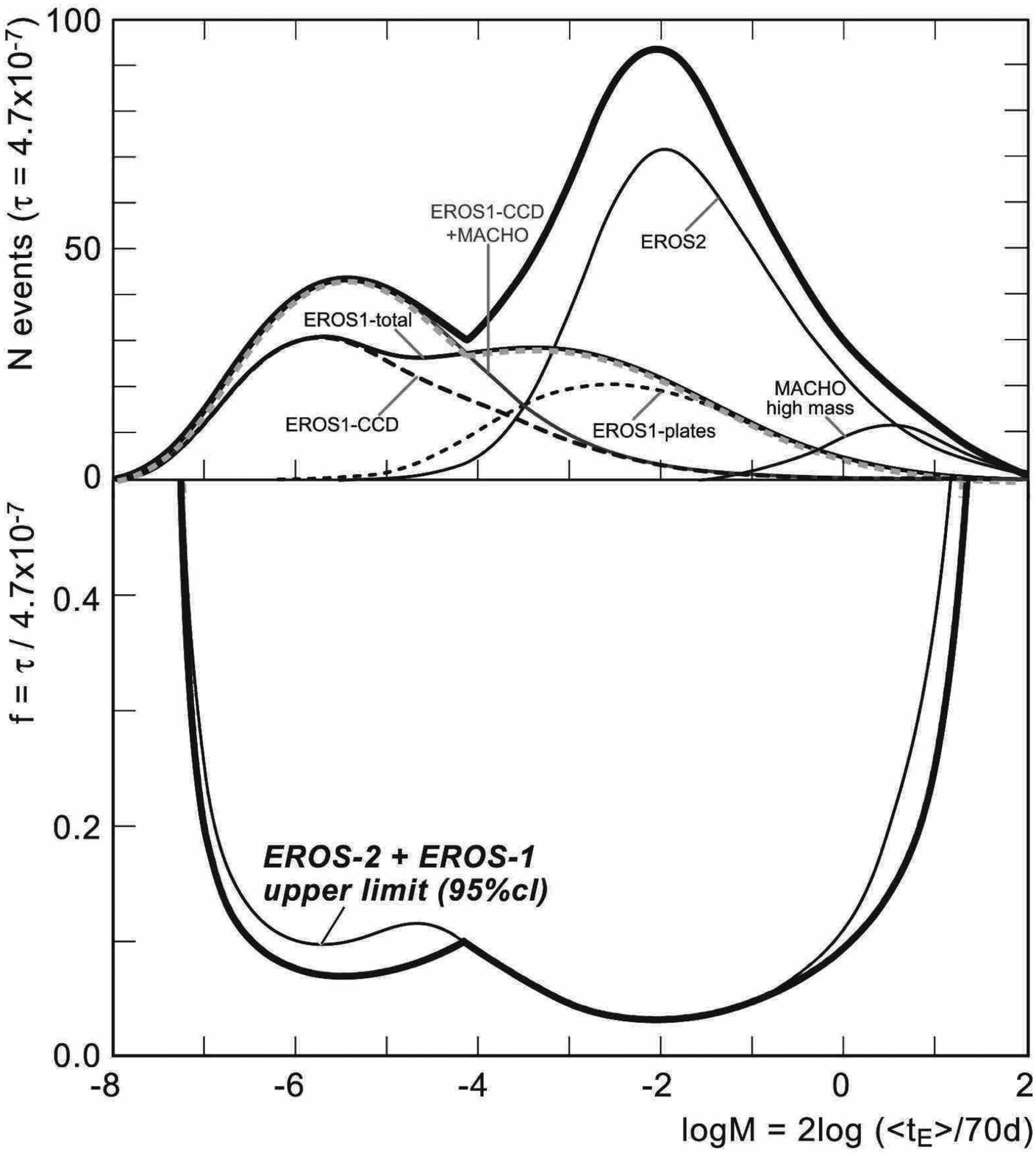}
\caption{\it
The top panel shows the number of expected events
towards LMC as a function of the lens mass $M$ for the S-model.
The EROS1-total line shows the sum of the expectations from the
CCD-camera and from the plates (dashed lines). The MACHO-high mass line 
corresponds to the ``zero event'' high mass search.
The EROS1-CCD+MACHO line (blue) is the combined result from
the EROS1 CCD-camera and the MACHO 2 year analysis.
The grey dashed line is the envelope of the EROS1-total and the
EROS1-CCD+MACHO lines.
The final combination (thick line) is the sum of EROS2,
of the grey dashed line and of half of the MACHO-high mass expectations (see text).
In the lower panel, the thick line shows the combined
$95\%$ CL upper limit on $f=\tau_{LMC}/4.7\times 10^{-7}$
based on no observed events.
The thin line is the EROS1+EROS2 LMC limit.
\label{combinaison}}
\end{center}
\end{figure}

The upper limit, considered as a function $f_{\delta}(M)$, is
the maximum contribution of objects of mass $M$ to the S-model halo optical depth,
expressed in fraction $f$.
The maximum contribution $f_F$ of a population with a (normalized)
mass distribution $F(m)$ different than a Dirac function $\delta(m-M)$,
is given by
\begin{equation}
f_F=\left[\int \frac{F(m)}{f_{\delta}(M)(m)}\ud m \right]^{-1}.
\end{equation}
We can combine the ``zero event'' analysis to obtain a stronger constraint
towards LMC. Including the non-zero event measurements from the SMC or from
the OGLE-II \cite{ogletauLMC} analysis in such a combination would need
a specific collaborative work,
since the corresponding exclusion limits depend on the expected
background of non-halo events in each survey (see below).
\vspace{-2mm}
\paragraph{\bf Combination of the limits on the low mass side:}
The simplest way to combine the $95\%$ exclusion results is to
consider the lower envelope of the exclusion curves
from all the null analyses.
A more sophisticated technique would be to calculate the exclusion
limits by adding the signals expected by each analysis,
taking into account only once those
objects that are monitored simultaneously by several surveys.
Figure \ref{combinaison} (top panel) shows the numbers of events expected from
the S-model halo lenses for each of the ``zero event'' experiments
(EROS1-CCD, EROS1-plates, EROS2, EROS1-CCD+MACHO, MACHO high-mass)
\footnote{in these experiments, the zero event observed are the very
final numbers after rejection from complementary observations, as
discussed just before this sub-section.}.
A complete combination, following \cite{EROSMACHO}
would need a close collaboration between the teams involved.
Nevertheless, a simple combination can be made:
\begin{itemize}
\item
let $n_{EROS1-plates}(M)$, $n_{EROS1-CCD+MACHO}(M)$
and $n_{EROS2}(M)$ be the expected numbers of events
as a function of the lens mass $M$ for the
corresponding analysis;
\item
since the EROS1-plates data overlap with the MACHO-2year observations,
they overlap with the combined EROS1-CCD+MACHO analysis.
Therefore, the total number of expected events from the EROS1-plates
plus the EROS1-CCD+MACHO analysis is not the sum of the expected numbers
but it is at least
larger than $n_{EROS1-plates}(M)$ and $n_{EROS1-CCD+MACHO}(M)$
(grey dashed line in Fig. \ref{combinaison}-top panel);
\item
since the data sets of the EROS1-plates and of the
combined EROS1-CCD+MACHO analysis do not overlap with
the EROS2 analysis the function
\begin{equation}
n_{EROS2}(M)+max[n_{EROS1-plates}(M),n_{EROS1-CCD+MACHO}(M)]
\end{equation}
represents a minimum for
the total number of events expected from all these analyses
(thick line in Fig. \ref{combinaison}-top panel).
\end{itemize}
Using this combination
allows us to improve the exclusion limit between $10^{-5}$ and $10^{-4}\Msol$
with respect to the envelope of the EROS/MACHO upper limits
(Fig. \ref{combinaison} (bottom panel)).

\vspace{-2mm}
\paragraph{\bf Combination of the limits on the high mass side:}
Another combination can be made on the high mass side
with the MACHO high-mass ``zero event'' analysis \cite{MACHOhighmass} and the
EROS results \cite{ErosLMCfinal};
the MACHO search did not overlap with the EROS 
searches for more than 3 years (1994-96), in the middle of the MACHO
$5.7$ year analysed observations; therefore one can conservatively
add half of the total published expectations from the
MACHO high-mass search
to the expectations from the other ``zero-signal'' searches.
This is conservative as the detection
efficiency ---specifically for long duration microlensing--- is better
for events that have their maximum
in the middle of the observation period.
%
\vspace{-2mm}
\paragraph{\bf The OGLE upper limit:}
No simple combination can be made with the OGLE-II results \cite{ogletauLMC},
as their field and observation epoch are entirely covered by the EROS2 program.
As for MACHO, the OGLE-II catalog of monitored stars contains fainter
stars 
than the EROS2 catalog, and a combination of these results
could potentially improve the limits but
would need a close collaboration between the teams.
Such a combination would also be much more complicated,
as the OGLE analysis is not a ``zero-signal'' analysis;
like the one SMC-event found by EROS and MACHO 
\cite{SMC2ans} \cite{MACHOSMC}, the 2 events
found by OGLE-II cannot be unambiguously affected to self-lensing.
Because of the existence of these events, according to standard statistics,
the OGLE-II limit should not be estimated 
by considering that 3 events correspond to the $95\%$ CL excluded mean value;
indeed, the $95\%$ CL upper limit on the mean signal $\bar{S}$ contaminated
by a background known by its expected mean value $\bar{B}$,
were $S+B$ follows a Poisson law, is not 3 when the observed number
is equal to $\bar{B}$.
In the present case, if one assumes that $\bar{B}=2$ (which is probably
a maximum according to \cite{ogletauLMC} \cite{Gyuk} \cite{macho2000LMC} 
\cite{belokurov2} and \cite{mancini2004}), then
following \cite{feldman}, the $95\%$ upper limit when $2$ events are
observed is $\bar{S} < 4.72$. This limit would be lower only if $\bar{B} > 2$.
If $\bar{B}=2$, the OGLE upper limit on the halo component should
be corrected by a factor $4.72/3=1.57$, but this needs to be finalized using
the exact value of $\bar{B}$.
The statistical procedure described in \cite{Eros1plaqlim}
could also be used for the OGLE-II non-zero event analysis.
\subsection{The Galactic halo}
\subsubsection{The apparent discrepancies towards the LMC}
We can quantify the incompatibility between the MACHO results from \cite{MACHO2yrs}
and \cite{macho2000LMC} and the EROS results \cite{ErosLMCfinal}.
The 13 observed events by MACHO in \cite{macho2000LMC}
correspond to a fraction $f_{MACHO}=0.21$ of the S-model halo;
The MACHO collaboration has estimated that
at least $9.6$ of these events should be due to
hidden baryonic matter (Milky-Way or LMC halo).
This corresponds to $9.6/f_{MACHO}=45.7$ events expected
with a full S-model halo entirely made of objects of mass $0.6\Msol$.
For the same mass, about $39\times f_{MACHO}=8.2$ events are expected in the
EROS1+EROS2 LMC data while zero were observed.
For a mass $0.9\Msol$, which is $1\sigma$ larger than the
central MACHO value, $30\times f_{MACHO}=6.3$ events were expected in the EROS data.
Assuming an optical depth dominated by the Galactic halo
and no correlation between the data sets,
the EROS and MACHO measurements are statistically incompatible.
The correlations due to the (actually small) overlap between the data sets makes
this incompatibility worse.

There are considerable differences between the EROS, MACHO and OGLE
data sets that may ---at least partially--- solve the apparent conflict.
Generally speaking, MACHO uses faint stars in dense fields
($1.2\times 10^7$ stars over $14\;deg.^2$) while EROS2 uses bright stars in
sparse fields ($0.7\times 10^7$ stars over $90\;deg.^2$).
As a consequence of the use of faint stars, only two of the 17 MACHO (set B)
candidates were sufficiently bright to be compared to the EROS bright sample;
as the corresponding events occurred before the EROS data taking, there is no
evidence for a malfunctionning experimental setup during one of the surveys.
The use of dense fields by the MACHO group also suggests that the
higher MACHO optical depth may be due, in part, to self-lensing in
the inner parts of the LMC \cite{Gyuk}.
The contamination of irregular variable objects faking microlensing in
low photometric resolution events could also be stronger in
the sample of faint stars used by MACHO.
As already mentioned, another difficulty when using faint source stars
originates in 
the blending effects that complicate the optical depth calculations.
\subsubsection{The SMC case}
In the framework of the S-model halo, the EROS upper limit on the halo fraction
from the combined LMC/SMC studies in \cite{SMC5ans} and \cite{ErosLMCfinal} assumes that
$\tau_{SMC-halo}=1.4\times \tau_{LMC-halo}$; this would be larger with a flattened halo
or a thick disk.
The dotted line in Fig. \ref{resultatsLMC}
shows the small impact of the SMC measurements according to this model.
Even though the impact on the optical depth estimates seems limited,
the compatibility of the SMC microlensing events population with the LMC one
is the subject of an interesting debate.
The long durations
of the four SMC events published
in the 5-year search of EROS \cite{SMC5ans} is found to be
incompatible with the $t_{\mathrm{E}}$ distribution of the MACHO LMC events \cite{macho2000LMC},
and make these events more compatible with unidentified variable-type stars
or with SMC-self-lensing. This issue is discussed by \cite{spectroSMC}.

Because of their long duration, a fair fraction of the SMC events could benefit
from complementary studies, thus providing indications on the lens population.
The first event towards SMC (MACHO-SMC-97-1/EROS-SMC1) was discovered in 1997 and observed
by both MACHO \cite{MACHOSMC} and EROS \cite{SMC1an}.
Thanks to its long duration ($t_{\mathrm{E}}=123\, days$) and its brightness,
a parallax analysis was performed on the light-curve, and from
the absence of measurable distorsion, it was concluded that the lens
probably belonged to the SMC itself.
This interpretation was confirmed from the Spectroscopy of the lensed star \cite{spectroSMC}.
The second event discovered towards SMC (MACHO-SMC-98-1)
was the first event from the Magellanic Clouds to benefit from an intensive
international collaboration between MACHO \cite{Machocaustic-SMC},
EROS \cite{EROS-causticSMC}, PLANET 
\cite{PLANET-caustic-SMC},
OGLE \cite{OGLE-causticSMC} and MPS\footnote{Microlensing Planet Search} 
\cite{MPS-causticSMC} collaborations on 18 June 1998,
as it was early detected by MACHO as a binary-lens microlensing event.
The common conclusion from the different survey teams was that the proper
motion of the lens was consistent
with the lens beeing within the SMC rather than in the Galactic plane or halo.
But at least one SMC-event, OGLE-2005-SMC-001, measured both from Earth and from
space \cite{space-SMC}, had a parallax estimate such that the
probability for the lens to
belong to the halo was 20 times larger than to belong to the SMC.
The two later events do not enter in any optical depth calculation, and
they are only mentioned here to illustrate the problem of the lens location.
Unfortunately, the very small statistics limits the interpretations,
but we can conclude that the SMC lens population is probably not dominated by the Galactic
halo, and is certainly different from the LMC lens population.
\subsubsection{A synthesis attempt: halo versus local structures}
Taking all the observational results at their face value, we may conclude
that the initial hypothesis of an optical depth dominated by
the Galactic halo ---almost uniform through all the monitored LMC fields
and 1.4 times larger towards SMC---
is wrong, because it cannot explain the EROS-MACHO differences
nor the LMC-SMC differences.

As the results found by the teams are mutually compatible
towards the Galactic plane directions (Sect. \ref{sec:population}),
the falseness of this
hypothesis should now be favored rather than the eventuality of wrong measurements.

Abandoning the original microlensing survey paradigm
of a dominant halo allows one to
explain the variety of the observations
by the large differences in the monitored fields. Fig. \ref{fieldsLMC}
supports this view.
If ---as suggested by their small contribution to the halo---
one assumes that lenses do not trace the Galactic halo hidden matter,
then the apparent discrepancy between the surveys can be understood
by considering local structures inducing self-lensing;
such structures (populations of foreground lenses as
well as populations of background sources \cite{macho2000HST})
may be responsible for
the variability of the mean optical depth with the monitored fields.
Nevertheless, a convincing proof of the existence of such structures
is still to be provided.
One of the aims of the current and next generations of microlensing
surveys towards LMC (like SuperMACHO) is to improve the
knowledge of the lens populations, and specifically
their location.

For a pertinent synthesis, it seems that the total optical depth
---understood
as an {\it average} over the monitored fields--- is the
perennial observable that will allow the building of contour
maps of the massive compact object spatial distribution.
The following numbers have to be {\it imperatively} associated with
the corresponding fields:
\begin{itemize}
\item
$\tau_{LMC-MACHO} = (1.0\pm 0.3)\times 10^{-7}$,
after revision by \cite{Bennett}.
There have been some controversy on the MACHO optical depth estimates.
The first estimate from \cite{MACHO2yrs} was clearly much too high
($\tau=2.9^{+1.4}_{-0.9}\times 10^{-7}$). This result was seriously
downsized to $\tau=1.2^{+0.4}_{-0.3}\times 10^{-7}$ after the
analysis of 5.7 years of data \cite{macho2000LMC}.
MACHO collaborators and outsiders have revised/disputed the
later result;
\cite{Bennett} took into account known contamination of the signal by
non-microlensing variabilities 
and published a revised optical depth of $\tau = (1.0\pm 0.3)\times 10^{-7}$.
\cite{belokurov1} and \cite{belokurov2} have re-analized the
publicly available MACHO light-curves with a neural network system and
estimated a number of contaminants larger than the one quoted
in the original MACHO
paper. They conclude that $\tau_{LMC-MACHO}=(0.4\pm 0.1)\times 10^{-7}$.
A controversy followed (\cite{Griest2005} and \cite{puzzle}).
\item
$\tau_{LMC-EROS} < 0.36\times 10^{-7}$ ($95\%$ CL).
\item
$\tau_{LMC-OGLE} = (0.43\pm 0.33)\times 10^{-7}$ is the total measured
optical depth (taking into account 2 observed events).
\end{itemize}
The distributions of event durations (except for EROS which has no event)
have also been used by several authors to refine the interpretations.
For example, a detailed study of the consequences of the MACHO results
was performed in \cite{Jetzer2002}, which takes into account
the optical depths, rates
and durations expected from several Galactic structures (thick disk, spheroid)
through the mass moment method \cite{Derujula91} \cite{Derujula92},
and concluded that at least
3 of the observed microlensing events were due to LMC lenses,
but that the others were probably due to other Galactic structures.
More recently \cite{LMC-Novati1} concluded
that up to about
half of the MACHO events could be attributed to the LMC dark matter halo
instead of the Milky-Way dark halo.
Calchi Novati {\it et al.} \cite{LMC-Novati2} also conclude from their rate
and durations that the 2 events reported by OGLE-II
are perfectly compatible with the expectation from known lenses
({\it i.e.} not dark), dominated by the LMC self-lensing.

{\bf Conclusion}

A substantial contribution of compact objects to a standard halo
is now clearly excluded. Compact objects belonging to
a flattened halo or a thick disk are also seriously limited
by observations both towards SMC and GSA.
Wether or not there is a small dark matter component under the form of
compact objects is still an open question.
%
\section{The future of the optical depth studies}
Several authors have investigated the possibility to
systematically use complementary observations
---instead of occasionnally as currently done---
to look for non-standard features.
Systematic searches
for parallax effects towards LMC have been examined by \cite{Rahvar-parallax}
based on the use of the ESO-NTT telescope, and by \cite{Sumi-parallax}
using measurements by the ESO-VLT or by the Hubble Space Telescope
on strongly magnified sources.
Such follow up measurements would increase the information about the events
and help to localize the lens populations.
Accurate mapping of the optical depth is another way to improve the knowledge
on these populations and to measure the self-lensing contributions.
Microlensing searches that address this question
are currently running or about to observe:
SuperMACHO \cite{supermacho1}, OGLE-III and IV \cite{OGLE3-realtime}
and the infrared VVV project \cite{VVVproposal}
using VISTA facility, should renew this scientific domain within the next
few years.
The LSST program will certainly dominate the following decade of
microlensing studies as a microlensing factory,
possibly with contributions from spatial missions like GAIA.
%
%
%
%
\begin{acknowledgements}
I am grateful to Jacques
Ha{\"\i}ssinski for his carefuly reading of the manuscript,
to Bruno Mazoyer for the production of the figures and to
Sebastiano Calchi Novati for his useful remarks and suggestions.
\end{acknowledgements}
%
%
%
%
%

\end{document}